\documentclass[conference,10pt]{IEEEtran}
\usepackage{balance}
\usepackage{graphicx}
\usepackage{bm}
\usepackage{amsmath}
\usepackage{amsfonts}
\usepackage{amssymb}
\usepackage{color}
\usepackage{mathrsfs}
\usepackage{upgreek}
\usepackage[utf8]{inputenc}
\usepackage{tikz}
\usetikzlibrary{arrows, calc, decorations.markings, positioning}
\usepackage{cite}
\usepackage{booktabs}
\usepackage{multirow}
\usepackage{amsmath}
\usepackage{algorithm}
\usepackage{algorithmic}
\usepackage{makecell}
\usepackage[flushleft]{threeparttable}
\usepackage{tabularx}
\usepackage{threeparttable}
 \usepackage[caption=false,font=footnotesize]{subfig}
\usepackage{amssymb}
\usepackage{amsmath}
\usepackage{overpic}
\usepackage{algorithm}
\usepackage{algorithmic}
\usepackage{multirow}
\usepackage{array}
\usepackage{graphicx}
\usepackage{bm}
\usepackage{color}
\usepackage{tikz}
\usepackage{epstopdf}
\usepackage{stfloats}
\usepackage{cite}
\usepackage{tikz}
\usepackage{amsmath}
\usepackage{psfrag}
\usepackage{acronym}
\usepackage{enumerate}
\usetikzlibrary{arrows}
\usetikzlibrary{decorations}

\definecolor{BLUE}{rgb}{0,0,1}

\newtheorem{corollary}{Corollary}
\newtheorem{proposition}{Proposition}
\newtheorem{remark}{Remark}
\newtheorem{lemma}{Lemma}

\acrodef{aoa}[AOA]{angle-of-arrival}
\acrodef{bcrb}[BCRB]{Bayesian Cram\'{e}r-Rao bound}
\acrodef{bfim}[BFIM]{Bayesian Fisher information matrix}
\acrodef{bp}[BP]{belief propagation}
\acrodef{cdi}[CDI]{cooperative dilution intensity}
\acrodef{cir}[CIR]{channel impulse response}
\acrodef{cl}[CL]{cooperative localization}
\acrodef{cp}[CP]{cyclic prefix}
\acrodef{crb}[CRB]{Cram\'{e}r-Rao bound}
\acrodef{crlb}[CRLB]{Cram\'{e}r-Rao lower bound}
\acrodef{dft}[DFT]{discrete Fourier transform}
\acrodef{dof}[DoF]{degree of freedom}
\acrodef{dpeb}[DPEB]{directional position error bound}
\acrodef{fim}[FIM]{Fisher information matrix}
\acrodef{efim}[EFIM]{equivalent Fisher information matrix}
\acrodef{hogs}[HOGS]{hybrid orthogonal-Gaussian signalling}
\acrodef{hsug}[HSUG]{hybrid semi-unitary--Gaussian}
\acrodef{ici}[ICI]{information coupling intensity}
\acrodef{icrb}[ICRB]{inverse CRB}
\acrodef{iid}[i.i.d.]{independently and identically distributed}
\acrodef{im}[IM]{index modulation}
\acrodef{isac}[ISAC]{integrated sensing and communication}
\acrodef{los}[LoS]{line-of-sight}
\acrodef{mse}[MSE]{mean-squared error}
\acrodef{ofdm}[OFDM]{orthogonal frequency-division multiplexing}
\acrodef{pdf}[PDF]{probability density function}
\acrodef{peb}[PEB]{position error bound}
\acrodef{speb}[SPEB]{squared position error bound}
\acrodef{pll}[PLL]{phase-locked loop}
\acrodef{psk}[PSK]{phase shift keying}
\acrodef{p2p}[P2P]{point-to-point}
\acrodef{qam}[QAM]{quadrature amplitude modulation}
\acrodef{rbs}[RBS]{reference broadcast synchronization}
\acrodef{rhs}[RHS]{right hand side}
\acrodef{rii}[RII]{ranging information intensity}
\acrodef{rss}[RSS]{received signal strength}
\acrodef{rc}[RC]{ranging coefficient}
\acrodef{speb}[SPEB]{squared position error bound}
\acrodef{toa}[TOA]{time-of-arrival}
\acrodef{tdoa}[TDOA]{time-difference-of-arrival}
\acrodef{tpsn}[TPSN]{time synchronization protocol for sensor network}
\acrodef{vmp}[VMP]{variational message passing}
\acrodef{wsn}[WSN]{wireless sensor network}
\acrodef{efim}[EFIM]{equivalent Fisher information matrix}
\acrodef{dio}[DIO]{distance-information-only}
\acrodef{aio}[AIO]{angle-information-only}
\acrodef{saaf}[SAAF]{squared array aperture function}
\acrodef{snc}[S\&C]{sensing and communications}
\acrodef{uoa}[UOA]{uniformly oriented array}
\acrodef{rgg}[RGG]{random geometric graph}
\acrodef{snr}[SNR]{signal-to-noise ratio}
\acrodef{eoc}[EoC]{efficiency of cooperation}
\acrodef{npi}[NPI]{nominal position information}
\acrodef{gnss}[GNSS]{global navigation satellite system}
\acrodef{mimo}[MIMO]{multiple-input multiple-output}
\acrodef{mcs}[MCS]{minimally constrained system}
\acrodef{zzb}[ZZB]{Ziv-Zakai lower bound}
\acrodef{wwb}[WWB]{Weiss-Weinstein lower bound}
\acrodef{nlos}[NLOS]{non-light-of-sight}
\acrodef{mmse}[MMSE]{minimum mean squared error}
\acrodef{uav}[UAV]{unmanned aerial vehicle}
\acrodef{ppp}[PPP]{Poisson point process}
\acrodef{bpp}[BPP]{binomial point process}
\acrodef{cln}[CLN]{cooperative location-aware network}
\acrodef{pdr}[PDR]{pedestrian dead reckoning}
\acrodef{ml}[ML]{maximum likelihood}
\acrodef{map}[MAP]{maximum \textit{a posteriori}}
\acrodef{kkt}[KKT]{Karush-Kuhn-Tucker}
\acrodef{st}[ST]{subspace tradeoff}
\acrodef{drt}[DRT]{deterministic-random tradeoff}
\acrodef{ustm}[USTM]{unitary space-time modulation}
\acrodef{pri}[PRI]{pulse repetition interval}
\acrodef{prf}[PRF]{pulse repetition frequency}
\acrodef{acf}[ACF]{autocorrelation function}
\acrodef{isl}[ISL]{integrated sidelobe level}
\acrodef{islr}[ISLR]{integrated sidelobe level ratio}
\acrodef{rgi}[MFI]{mainlobe fluctuation intensity}
\acrodef{irgi}[AMFI]{averaged mainlobe fluctuation intensity}
\acrodef{cds}[CDS]{cyclic difference set}
\acrodef{psl}[PSL]{peak sidelobe level}
\acrodef{aesl}[AESL]{average expected sidelobe level}
\acrodef{pesl}[PESL]{peak expected sidelobe level}
\acrodef{masm}[MASM]{MASked Modulation}
\acrodef{si}[SI]{self-interference}
\acrodef{cpi}[CPI]{coherent processing interval}

\acrodefplural{dof}[DoFs]{degrees of freedom}
\acrodefplural{drt}[DRTs]{deterministic-random tradeoffs}
\acrodefplural{cds}[CDSs]{cyclic difference sets}

\usepackage{winsnotation}
\title{Transmission Mask Analysis for Range–Doppler Sensing in Half-Duplex ISAC}
\author{Dikai Liu$^\dag$, Yifeng Xiong$^\dag$, Marco Lops$^\ddag$, Fan Liu$^\natural$, and Jianhua Zhang$^\sharp$\\
\small$^\dag$School of Information and Communication Engineering, Beijing University of Posts and Telecommunications, China\\
$^\ddag$Department of Electrical and Information Technology, University of Naples Federico II, Italy\\
$^\natural$National Mobile Communications Research Laboratory, Southeast University, China
\\
$^\sharp$State Key Laboratory of Networking and Switching Technology, Beijing University of Posts and Telecommunications, China\\
Email: ldk@bupt.edu.cn, yifengxiong@bupt.edu.cn, lops@unina.it, fan.liu@seu.edu.cn, jhzhang@bupt.edu.cn

}

\begin{document}
\maketitle

\begin{abstract}
In this paper, we analyze the periodic transmission masks for \acf{masm} in half-duplex \acf{isac}, and derive their closed-form expected range-Doppler response $\mathbb{E}\{|r(k,l,\nu)|^2\}$. We show that range sidelobes ($k\neq l$) are Doppler-invariant, extending the range-sidelobe optimality to the 2-D setting. For the range mainlobe ($k=l$), periodic masking yields sparse Doppler sidelobes: \Acp{cds} (in particular Singer \acp{cds}) are minimax-optimal in a moderately dynamic regime, while in a highly dynamic regime the Doppler-sidelobe energy is a concave function of the mask autocorrelation, revealing an inevitable tradeoff with mainlobe fluctuation.
\end{abstract}


\section{Introduction}
\Acf{isac} aims to provide sensing and communication services over a shared radio platform, promising substantial gains in spectral/energy efficiency and hardware reuse \cite{chafii2023twelve,saad2019vision,XiongTIT,cui2021integrating,li2022integrated,liu2022integrated}. Among various design paradigms, a communication-centric route is particularly appealing, as it leverages mature communication waveforms and protocols with minimal changes \cite{sturm2011waveform,chen2021code,zeng2020joint,gaudio2020effectiveness,yuan2024otfs,liu2025cp}. However, straightforward sensing with continuous transmission faces a fundamental bottleneck: insufficient transmitter-receiver isolation leads to strong \ac{si} \cite{sabharwal2014band}, which can severely limit the sensing dynamic range and thus the achievable sensing range. A classical remedy in radar systems is half-duplex operation \cite{wang2023bidirectional}, namely, switching off the receiver during transmission, as in pulse radars. Conventional ranging-oriented pulse radars typically employ narrow pulses and small duty cycles, which conflict with the high-throughput requirement in \ac{isac}.

To bridge this tension, \acf{masm} \cite{11165031} was proposed as a high-duty-cycle half-duplex \ac{isac} waveform design framework. The key idea is to multiply a data-bearing symbol stream by a periodic binary transmission mask, and to apply a corresponding reception mask at the receiver, thereby enabling \ac{si}-free operation while still carrying information-bearing symbols over a large fraction of time. The original contribution of \ac{masm} focused on the range dimension and introduced the notion of \emph{mainlobe fluctuation} to quantify the unevenness of the range-compression mainlobe across range bins induced by half-duplex masking. Importantly, Doppler effects were not considered therein, and the analysis was restricted to the range response.

Nevertheless, in many \ac{isac} applications, targets are moving, and a \emph{range-Doppler} characterization becomes indispensable. When coherent processing spans multiple \acp{pri}, Doppler induces phase rotations across fast-time samples and across \acp{pri}, and the interaction between Doppler and the \emph{periodic} masking pattern may produce structured Doppler sidelobes and ambiguity phenomena. This motivates a principled extension of \ac{masm} analysis and mask design from the range axis to the full range-Doppler plane.

In this paper, we study the range-Doppler response of \ac{masm} waveforms through its expected squared magnitude $\mathbb{E}\{|r(k,l,\nu)|^2\}$, which subsumes: (i) the range-Doppler mainlobe at the true range bin, (ii) Doppler sidelobes on the range mainlobe, (iii) range sidelobes on the Doppler mainlobe, and (iv) generic sidelobes. We further distinguish two Doppler regimes. In the \emph{moderately dynamic} regime, Doppler-induced phase rotation within a single \ac{pri} is limited, so range compression can be performed \ac{pri}-by-\ac{pri}, followed by Doppler estimation across \acp{pri}. In the \emph{highly dynamic} regime, range compression and Doppler estimation are no longer decoupled, and ambiguity effects become more prominent.

Our main findings can be summarized as follows:
\begin{itemize}
  \item \textbf{Range sidelobes are Doppler invariant.} For $k\neq l$, we show that $\mathbb{E}\{|r(k,l,\nu)|^2\}$ is independent of the Doppler index $\nu$. Consequently, classical mask optimality results for peak range sidelobes extend naturally to the range-Doppler setting.

  \item \textbf{\Acfp{cds} are optimal masks in the moderately dynamic regime.} For $k=l$, we characterize the Doppler-sidelobe structure induced by periodic masking and identify a strong sparsity property of Doppler sidelobes, implying that maximizing the off-zero periodic autocorrelation floor of the mask would simultaneously suppress the worst Doppler sidelobes and reduces mainlobe fluctuation. \acp{cds}, in particular Singer \acp{cds}, emerge as optimal masks in this regime under a natural minimax criterion.

  \item \textbf{Ideal masks do not exist in the highly dynamic regime.} When Doppler exceeds the (conventional) unambiguous region, strictly periodic pulse trains exhibit grating lobes at Doppler locations spaced by the \ac{prf}. We quantify the aggregate deterministic Doppler sidelobe energy along the range mainlobe and show that it admits a concise expression as a concave function of the mask autocorrelation. This reveals that the mask patterns that minimize mainlobe fluctuation cannot, in general, simultaneously minimize the Doppler sidelobes in the highly dynamic regime.
\end{itemize}

The above results offer explicit guidance on when \ac{cds}-type masks remain ideal and when tradeoffs are inevitable. The remainder of the paper is organized as follows. Section~II introduces the signal model as well as the range-Doppler response metric. Section~III analyzes $\mathbb{E}\{|r(k,l,\nu)|^2\}$ and establishes the main results in the two Doppler regimes. Section~IV demonstrates the analytical results using numerical examples. Section~V concludes the paper.

\section{System Model}
\subsection{Transmission and Sensing Reception Model}
For \ac{masm} systems using pulse shapers satisfying the Nyquist criterion, the transmitted signal can be discretized into the following sequence \cite{11165031}
$$
x[n] = m_{\rm t}[n]x_n,
$$
where $m_{\rm t}[n]$ denotes the transmission mask which is $N$-periodic and $0/1$-valued, and $\{x_i\}$ denotes a zero-padded data sequence with $x_i=0$ when $m_{\rm t}[i]=0$, and $x_i\in\Set{S}$ when $m_{\rm t}[i]=1$, where $\Set{S}$ denotes the constellation. Each period of $m_{\rm t}[n]$, namely $N$, is referred to as a \ac{pri}, and we consider a \ac{cpi} consisting of $M$ \acf{pri}. We assume that all constellation points are selected with equal probabilities during the modulation processing, and that
$$
\begin{aligned}
&\mathbb{E}\{x_i|m_{\rm t}[i]=1\} = 0,~\mathbb{E}\{x_i^2|m_{\rm t}[i]=1\} = 0, \\
&\mathbb{E}\{|x_i|^2|m_{\rm t}[i]=1\} = 1,~\mathbb{E}\{|x_i|^4|m_{\rm t}[i]=1\} = \mu_4,
\end{aligned}
$$
hold for all $x_i$'s. We also assume that all $x_i$'s with $m_{\rm t}[i]=1$ are mutually independent.

The received echo from a single target (neglecting the noise) is a delayed and Doppler-shifted version of $x[n]$, given by
$$
z_{k,\nu_0}[n] = m_{\rm r}[n]m_{\rm t}[n-k]x_{n-k}e^{\frac{j2\pi\nu_0n}{MN}},
$$
where $m_{\rm r}[n]=1-m_{\rm t}[n]$ denotes the reception mask, $k$ denotes the delay, and $\nu_0$ denotes the Doppler shift. Upon reception of the echo, the receiver performs a two-dimensional correlation, yielding the following three-dimensional range-Doppler response
$$
\begin{aligned}
r(k,l,\nu) &= \sum_{n=0}^{MN-1} z[n](m_{\rm t}[n-l]x_{n-l})^* e^{-\frac{j2\pi \nu_{\rm t} n}{MN}}\\
&=\sum_{n=0}^{MN-1} m_{\rm r}[n]m_{\rm t}[n-k]m_{\rm t}[n-l]x_{n-k}x_{n-l}^*e^{-\frac{j2\pi \nu n}{MN}},
\end{aligned}
$$
where $\nu=\nu_{\rm t}-\nu_0$, and $\nu_{\rm t}$ denotes the trial Doppler shift. We note that $r(k,l,\nu)$ depends on $k$ and $l$ as two independent variables due to the fact that half-duplex sensing incurs mainlobe fluctuation, while it depends on $\nu_{\rm t}$ and $\nu_0$ only via their difference $\nu$ since the masks do not filter out Doppler frequency shift components.

\subsection{Performance Metrics}
Qualitatively speaking, an ideal transmission mask should yield small mainlobe fluctuation and low sidelobe levels. This can be quantified using the following quantity
\begin{align}\label{fourth_all}
\mathbb{E}\{|r(k,l,\nu)|^2\} &= \sum_{n=0}^{MN-1}\sum_{m=0}^{MN-1}m_{\rm r}[n]m_{\rm r}[m]m_{\rm t}[n-k] \nonumber \\
&\hspace{3mm}\cdot m_{\rm t}[m-k]m_{\rm t}[n-l]m_{\rm t}[m-l]e^{\frac{j2\pi \nu (n-m)}{MN}}\nonumber \\
&\hspace{3mm} \cdot \mathbb{E}\{x_{n-k}x_{n-l}^*x_{m-k}^*x_{m-l}\}.
\end{align}
Depending on the specific indices, the physical meaning of this quantity is summarized as follows.
\begin{itemize}
    \item $k=l,~\nu=0$: The mainlobe of the range-Doppler response when the true range is in the $k$-th range bin.
    \item $k=l,~\nu\neq 0$: The Doppler sidelobes on the range mainlobe at the $k$-th range bin.
    \item $k\neq l,~\nu=0$: The Range sidelobes on the Doppler mainlobe $\nu=0$.
    \item Other configurations: Generic sidelobes.
\end{itemize}

In this paper, we consider a sensing range within a single \ac{pri}, in the sense that we only consider the region of $k,l\in\{1,2,\dotsc,N-1\}$ (note that $k=0$ or $l=0$ can be safely ignored since it is the blind range \cite{11165031}). We would wish to have the mainlobe levels $\mathbb{E}\{|r(k,k,0)|^2\}$ being constant over $k=1,2,\dotsc,N-1$, while keeping the sidelobe levels $\mathbb{E}\{|r(k,l,\nu)|^2\}$, where $k=l$ and $\nu=0$ do not simultaneously hold, as small as possible.

Regarding the value of Doppler frequency shifts, we identify two regimes:

\begin{enumerate}
\item \textbf{Moderately dynamic regime} ($\nu_0 < M$): In this regime, all targets have true Doppler shifts $\nu_0$ bounded by $M$. Note that $\nu$ is defined as the trial Doppler minus the true Doppler ($\nu = \nu_{\rm t} - \nu_0$). Under the condition $\nu_0 < M$ for all targets, we only need to consider $\mathbb{E}\{|r(k,l,\nu)|^2\}$ in the region $\nu < M$, since larger Doppler mismatches will not occur. In practice, this means range and Doppler processing can be decoupled\footnote{In terms of performance, this treatment works well only when $\nu_0\ll M$.}: one can first perform range compression within each \ac{pri}, and then estimate Doppler frequencies by applying spectral estimation across the sequence of range-compressed samples, with sampling rate equal to the \ac{prf}.
\item \textbf{Highly dynamic regime} ($\nu_0 \ge M$): If any target has a true Doppler $\nu_0$ exceeding $M$, the above separation no longer holds, necessitating joint range-Doppler processing. In this case, we would have to take into account $\mathbb{E}\{|r(k,l,\nu)|^2\}$ in the region $\nu \geq M$.
\end{enumerate}

\section{Analysis of the Range-Doppler Response}
In this section, we analyze $\mathbb{E}\{|r(k,l,\nu)|^2\}$. In particular, we only consider Doppler differences at $\nu=\{0,1,\dotsc,MN-1\}$. The characteristic at fractional Doppler differences can be adjusted using the windowing technique, and can be depicted using the ``iceberg--sea-level'' decomposition \cite{iceberg}.

\subsection{The range sidelobes $k\neq l$}\label{ssec:range_sidelobe}
Let us first consider the range sidelobes corresponding to $k\neq l$. Using the fact that symbols having different indices are mutually independent, we now have
\begin{equation}
\mathbb{E}\{x_{n-k}x_{n-l}^*x_{m-k}^*x_{m-l}\} = \left\{
\begin{array}{ll}
1, & \hbox{$n=m$;}\\
0, & \hbox{otherwise.}
\end{array}
\right.
\end{equation}
This implies that
\begin{equation}
\mathbb{E}\{|r(k,l,\nu)|^2\}= \sum_{n=0}^{MN-1} m_{\rm r}[n]m_{\rm t}[n-k]m_{\rm t}[n-l],
\end{equation}
which turns out to be independent of the Doppler difference $\nu$. Moreover, since $m_{\rm t}[n]$ and $m_{\rm r}[n]$ both have period $N$, we further have
\begin{equation}
\mathbb{E}\{|r(k,l,\nu)|^2\}= MR_{k,l},
\end{equation}
where $R_{k,l}=\sum_{n=0}^{N-1} m_{\rm r}[n]m_{\rm t}[n-k]m_{\rm t}[n-l]$. It has been shown in \cite{11165031} that the average range sidelobe levels for $\nu=0$ in a single \ac{pri}, namely
\begin{equation}\label{average_sidelobe}
\frac{\sum_{k=1}^{N-1} \sum_{l=1\atop l\neq k}^{N-1} R_{k,l}}{(N-1)(N-2)} = \frac{\rho(1-\rho)(\rho N-1)N^2}{{(N-1)(N-2)}},
\end{equation}
does not rely on the transmission mask $m_{\rm t}[n]$. Now, since $\mathbb{E}\{|r(k,l,\nu)|^2\}$ is independent of $\nu$, it now turns out that the sum of all range sidelobe levels
$$
\sum_{\nu=0}^{MN-1}\sum_{k=1}^{N-1}\sum_{l=1\atop l\neq k}^{N-1} \mathbb{E}\{|r(k,l,\nu)|^2\}
$$
would also be independent of transmission masks. Nevertheless, if we consider the peak sidelobe levels, namely
$$
\max_{k\neq l\atop k\neq 0,l\neq 0}~\mathbb{E}\{|r(k,l,0)|^2\},
$$
the optimal transmission masks would be those belonging to the class of Singer \ac{cds} \cite{gordon2015jolla,singer1938theorem}. Similarly, extending the result to the range-Doppler sensing scenario, we have
\begin{equation}
\max_{\nu} \max_{k\neq l\atop k\neq 0,l\neq 0}~\mathbb{E}\{|r(k,l,\nu)|^2\} = \max_{k\neq l\atop k\neq 0,l\neq 0}~\mathbb{E}\{|r(k,l,0)|^2\},
\end{equation}
and hence Singer \acp{cds} would also be ideal in the sense of peak sidelobe levels.

\subsection{The range mainlobes $k=l$, Moderately Dynamic Regime}\label{ssec:kl_moderate}
Now let us turn to the case of range mainlobe, namely when $k=l$. In this case, we have
$$
\mathbb{E}\{x_{n-k}x_{n-l}^*x_{m-k}^*x_{m-l}\} = \mathbb{E}\{|x_{n-k}|^2|x_{m-k}|^2\},
$$
which implies that
\begin{equation}\label{fourth_mainlobe}
\mathbb{E}\{x_{n-k}x_{n-l}^*x_{m-k}^*x_{m-l}\} = \left\{
\begin{array}{ll}
\mu_4, & \hbox{$m=n$;}\\
1, & \hbox{$m\neq n$,}
\end{array}
\right.
\end{equation}
since different symbols are mutually independent. Substituting \eqref{fourth_mainlobe} into \eqref{fourth_all}, we obtain
\begin{align}\label{kknu}
&\mathbb{E}\{|r(k,k,\nu)|^2\} \nonumber \\
&\hspace{3mm}= \mu_4 \sum_{n=0}^{MN-1} m_{\rm r}[n] m_{\rm t}[n-k] \nonumber \\
&\hspace{6mm} + \sum_{m\neq n} m_{\rm r}[n]m_{\rm r}[m]m_{\rm t}[n-k]m_{\rm t}[n-k] e^{\frac{j2\pi \nu(n-m)}{N}} \nonumber\\
&\hspace{3mm}= |S_{k,MN}(\nu)|^2+(\mu_4-1)M(\rho N-a[k]),
\end{align}
where
\begin{subequations}
\begin{align}
S_{k,MN}(\nu) &= \sum_{n=0}^{MN-1} m_{\rm r}[n]m_{\rm t}[n-k]e^{-\frac{j2\pi\nu n}{MN}},\\
a[k] &= \sum_{n=0}^{N-1}m_{\rm t}[n]m_{\rm t}[n-k].
\end{align}
\end{subequations}
In fact, the sequence $a[k]$ is the periodic autocorrelation sequence of the transmission mask. The term $|S_{k,MN}(\nu)|^2$ denotes the deterministic part of the ambiguity function mainlobe when $\nu=0$, while for $\nu\neq 0$ it represents the deterministic part of the sidelobe level along the Doppler axis. We observe that the term $|S_{k,MN}(\nu)|^2$ determines the shape of the Doppler sidelobes, since the term $(\mu-1)M(\rho N-a[k])$ is constant with respect to $\nu$. Next, we characterize an important property of $|S_{k,MN}(\nu)|^2$.
\begin{lemma}\label{lem:pulse_train}
For $\nu\in\{0,1,\dotsc,MN-1\}$, we have $S_{k,MN}(\nu)=0$ for $\nu\not\equiv 0\pmod{M}$.
\end{lemma}
\begin{IEEEproof}
The sequence $S_{k,MN}(\nu),~\nu=0,\dotsc MN-1$ is seen to be the \ac{dft} of $\gamma[n]:=m_{\rm r}[n]m_{\rm t}[n-k]$. Since both $m_{\rm t}[n]$ and $m_{\rm r}[n]$ have period of $N$, the sequence $\gamma[n]$ also has a period of $N$. This implies that its \ac{dft} only has non-zero values when $\nu$ is an integer multiple of $MN/N=M$, which completes the proof.
\end{IEEEproof}

From Lemma \ref{lem:pulse_train} we see that in the moderately dynamic regime ($\nu<M$), the range mainlobes satisfy
\begin{align}
&\mathbb{E}\{|r(k,k,\nu)|^2\} \nonumber \\
&\hspace{3mm}=\left\{
\begin{array}{ll}
    M^2(\rho N-a[k])^2+(\mu_4-1)M(\rho N-a[k]),~\nu=0; & \hbox{} \\
    (\mu_4-1)M(\rho N-a[k]),~\nu=1,\dotsc,M-1. & \hbox{}
\end{array}
\right.
\end{align}
Naturally, we would like to minimize the Doppler sidelobes corresponding to $\nu=1,\dotsc,M-1$. Since $\mu_4 \geq 1$, this is equivalent to maximizing $a[k]$. However, note that the sum of $a[k]$'s is constant:
$$
\begin{aligned}
\sum_{k=0}^{N-1}a[k]&=\sum_{n=0}^{N-1} m_{\rm t}[n] \sum_{k=0}^{N-1}m_{\rm t}[n-k] \\
&=\rho^2 N^2.
\end{aligned}
$$
It is thus more reasonable to minimize the maximal sidelobe, in the sense to solve
\begin{equation}\label{minimax_sidelobe}
\max_{m_{\rm t}[n]}\min_{k\neq 0}~a[k].
\end{equation}
The optimum of \eqref{minimax_sidelobe} is achieved when $a[k]$ is constant for all $k\neq 0$, which in turn is achievable by \acp{cds}. We may therefore arrive at the following conclusion.
\begin{remark}
The Singer \acp{cds} are the optimal transmission masks for range-Doppler sensing in the moderately dynamic regime, in the sense that it yields the lowest mainlobe fluctuation, and achieves the lowest maximal sidelobe levels (for $\nu\in\{0,\dotsc,M-1\}$ and $k,l\in\{0,\dotsc,N-1\}$).
\end{remark}

\subsection{The range mainlobes $k=l$, Highly Dynamic Regime}\label{ssec:kl_high}
Next let us consider the highly dynamic regime $\nu\geq M$. In this case, we could still rely on the expression \eqref{kknu}. According to Lemma \ref{lem:pulse_train}, we may focus on the locations at which $\nu = nM,~n=1,\dotsc N$, since otherwise $|S_{k,MN}(\nu)|^2=0$. In effect, we can simply consider $\mathbb{E}\{|k,l,nM|^2\}$, for which we may compute the term
\begin{subequations}
\begin{align}
S_{k,N}(\nu) &= \sum_{n=0}^{N-1}m_{\rm r}[n]m_{\rm t}[n-k]e^{-\frac{j2\pi\nu n}{N}} \\
&=\sum_{n=0}^{N-1}m_{\rm r}[n]m_{\rm t}[n-k]e^{-\frac{j2\pi\nu nM}{NM}} \\
&=\frac{1}{M}\sum_{n=0}^{MN-1}m_{\rm r}[n]m_{\rm t}[n-k]e^{-\frac{j2\pi\nu nM}{NM}} \label{one_over_m}\\
&=\frac{1}{M} S_{k,MN}(\nu),
\end{align}
\end{subequations}
where \eqref{one_over_m} follows from the fact that the sequence $e^{-\frac{j2\pi\nu nM}{NM}}$ has a period of $N$. Using the terminology from the radar literature, $\nu_{\max}=M-1$ is the \emph{maximal unambiguous Doppler} when ambiguity resolving techniques are not applied, since for strictly periodic pulse trains there would indeed be grating lobes at $\nu = nM,~n=1,\dotsc N$. In this context, we may say that \ac{masm}
is capable of resolving the Doppler ambiguity to a degree, and one would naturally wish to further enhance the ambiguity resolving capability by designing appropriate transmission masks.

Against this background, let us first investigate the sum of these Doppler sidelobes.
\begin{proposition}\label{prop:doppler_sidelobe_range_main}
The sum of the deterministic parts of sidelobe levels along the range mainlobe $k=l$ is given by
\begin{equation}\label{s_nu}
\sum_{\nu=1}^{N-1} |S_{k,N}(\nu)|^2 = (\rho N-a[k])[N-(\rho N-a[k])].
\end{equation}
\begin{IEEEproof}
Please refer to Appendix \ref{sec:proof_doppler_sidelobe_range_main}.
\end{IEEEproof}
\end{proposition}

Now, observe that $f(a[k])=\sum_{\nu=1}^{N-1} |S_{k,N}(\nu)|^2$ is a concave function of $a[k]$, which implies the following result.
\begin{proposition}\label{prop:eisl_max}
Given a fixed duty cycle $\rho$, we have
\begin{align}
I_{a[k]}:&=\sum_{n=1}^{N-1}\mathbb{E}\{|r(k,k,nM)|^2\}\nonumber \\
&\leq \rho(1\!-\!\rho) N^2\left(N\!-\!\frac{\rho(1\!-\!\rho) N^2}{N-1}+(N\!-\!1)(\mu_4\!-\!1)\right),
\end{align}
where the equality is achieved when all $a[k]$'s take identical values.
\begin{IEEEproof}
Please refer to Appendix \ref{sec:proof_eisl_max}.
\end{IEEEproof}
\end{proposition}

Another issue that raises concern is the mainlobe fluctuation phenomenon, a unique characteristic of half-duplex sensing schemes, which refers to the unevenness of $\mathbb{E}\{|r(k,k,0)|^2$ at different values of $k$. Since there is no Doppler difference in this case, we may directly apply the result in \cite{11165031}:
\begin{equation}\label{mainlobe_at_k}
\mathbb{E}\{|r(k,k,0)|^2 = (a[k])^2 + (\mu_4-1) (\rho N-a[k]),
\end{equation}
which achieves the minimal fluctuation when all $a[k]$'s ($k\neq 0$) are identical. It is then clear from Proposition \ref{prop:eisl_max} and \eqref{mainlobe_at_k} that there is a clash between mainlobe fluctuation and the sum of Doppler sidelobe levels, in the sense that they cannot be simultaneously minimized. To be specific, transmission masks forming \acp{cds} would yield identical $a[k]$ values for $k\neq 0$, which is ideal in the sense of mainlobe flctuation, but would at the same time have the worst sum of Doppler sidelobe levels, since they achieve the upper bound in Proposition \ref{prop:eisl_max}. On the other hand, other transmission masks containing highly polarized values of $a[k]$ would result in severe mainlobe fluctuation, but have smaller sums of Doppler sidelobe levels. Next we show what kind of masks yield the smallest sums of Doppler sidelobe levels, but as we shall see later, they are not practically favorable.
\begin{proposition}\label{prop:doppler_lower_bound}
The term $I_{a[k]}$ is bounded from below by
\begin{align}\label{doppler_lower_bound}
I_{a[k]} &\geq \rho (1-\rho )^2 N^3+(\mu_4-1) \rho(1-\rho)N^2(N-1).
\end{align}
When $d=1/\rho$ is an integer, the lower bound in \eqref{doppler_lower_bound} can be achieved by comb-like sequences taking the following form
\begin{equation}\label{comb_seq}
m_{\rm t}[n] = \left\{
\begin{array}{ll}
1, & \hbox{$n\equiv 0\pmod{(1/\rho)}$;} \\
0, & \hbox{otherwise,}
\end{array}
\right.
\end{equation}
or their cyclic-shifted versions.
\begin{IEEEproof}
Please refer to Apppendix \ref{sec:proof_doppler_lower_bound}.
\end{IEEEproof}
\end{proposition}

From a practical perspective, the comb-like sequences depicted by \eqref{comb_seq} are not favorable masks for long-range sensing, since they would introduce high sidelobes (nearly grating lobes at range bins whose indices are integer multiples of $d$. This suggests that additional performance metrics are in need. Another reasonable metric is the worst-case sum of Doppler sidelobe levels at range mainlobes, given by
$$
J_{a[k]} = \max_{k}~\sum_{n=1}^{N-1}\mathbb{E}\{|r(k,k,nM)|^2\},
$$
for which we have the following result.
\begin{corollary}\label{coro:pesl}
The transmission masks corresponding to \acp{cds} achieve the minimum $J_{a[k]}$ among all masks having the same values of $N$ and $\rho$.
\begin{IEEEproof}
Note that $f(a[k])$ decreases with respect to $a[k]$ when $a[k]\geq \left(\rho -\frac{1}{2}\right)N$, which implies that $f(a[k])+(\mu_4-1)(\rho N-a[k])$ is also a decreasing function of $a[k]$. Thus minimizing $J_{a[k]}$ amounts to maximizing $\min_k a[k]$. Under the constraint that $\sum_{k=1}^{N-1} a[k]=\rho N(\rho N-1)$ (see \eqref{from_zero_one}), we see that $\min_k a[k]$ is maximized when all values of $a[k]$ (except for $k=0$) are identical, which corresponds to the case of \acp{cds}. According to \cite{11165031}, $a[k]\geq \left(\rho -\frac{1}{2}\right)N$ satisfies for any $\rho\in(0,1)$, and hence the proof is completed.
\end{IEEEproof}
\end{corollary}

In light of Corollary \ref{coro:pesl}, we may still consider \acp{cds} as beneficial transmission masks for long-range sensing tasks in highly dynamic scenarios. For short-range sensing, one may resort to the comb-like sequences leading to lower Doppler sidelobes for $k < 1/\rho$.

\begin{figure*}[t]
  \centering
  \subfloat[Singer \ac{cds}]{
    \includegraphics[width=0.43\textwidth]{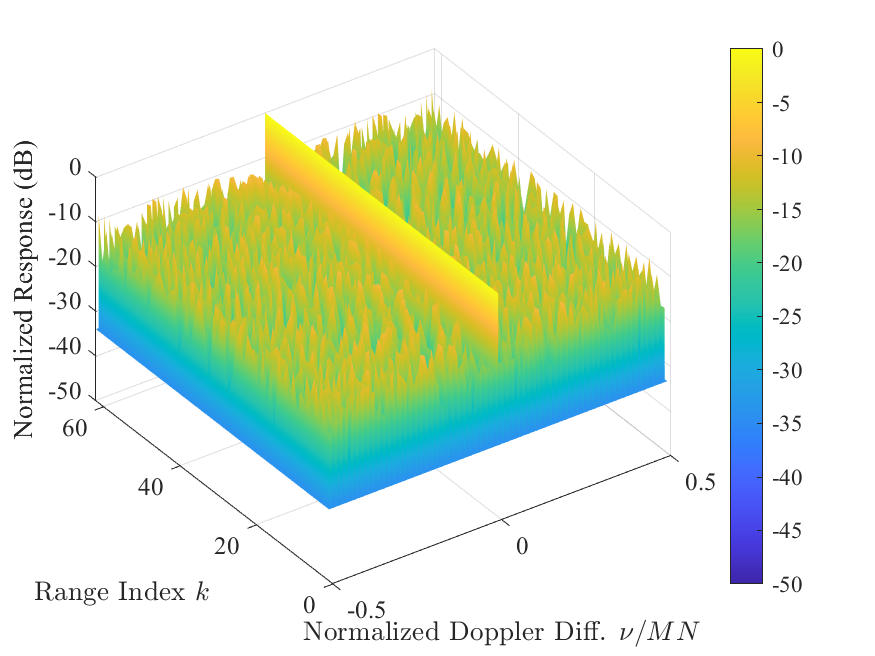}
    \label{fig:doppler_global_cds}
  }
  \subfloat[Random mask]{
    \includegraphics[width=0.43\textwidth]{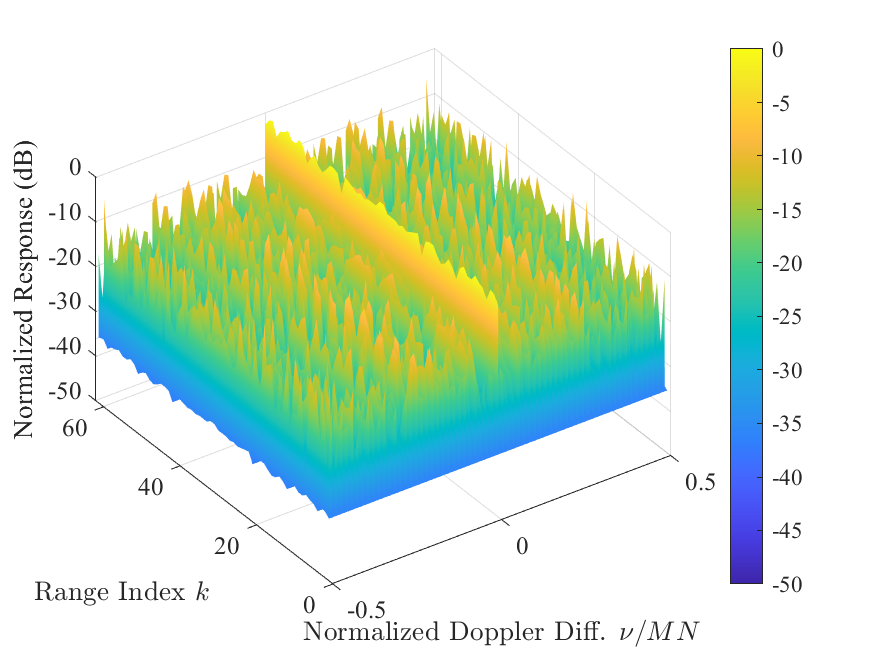}
    \label{fig:doppler_global_rand}
  }
  \\
  \subfloat[Comb-like mask]{
    \includegraphics[width=0.43\textwidth]{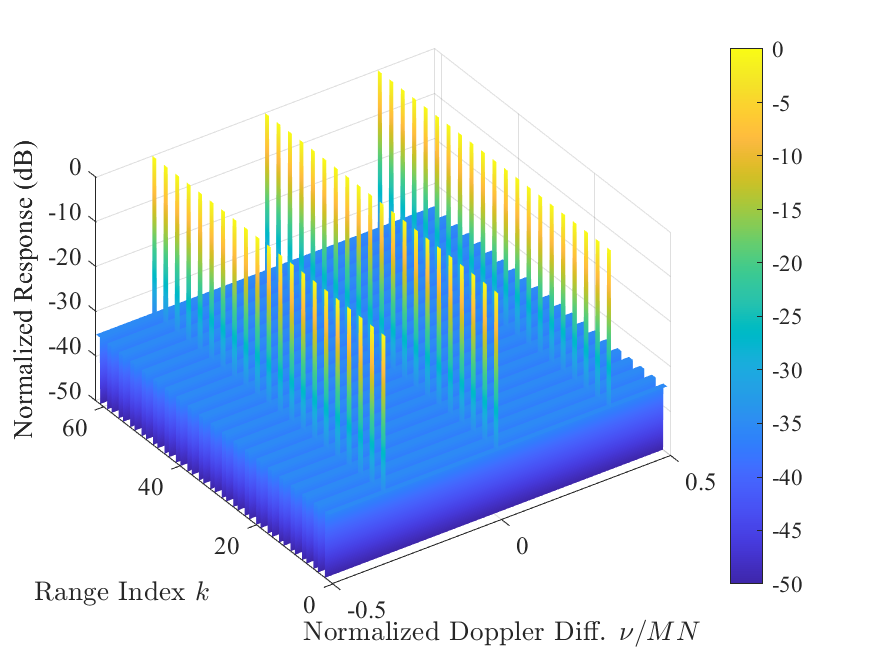}
    \label{fig:doppler_global_comb}
  }
  \subfloat[Mean Doppler sidelobe]{
    \includegraphics[width=0.43\textwidth]{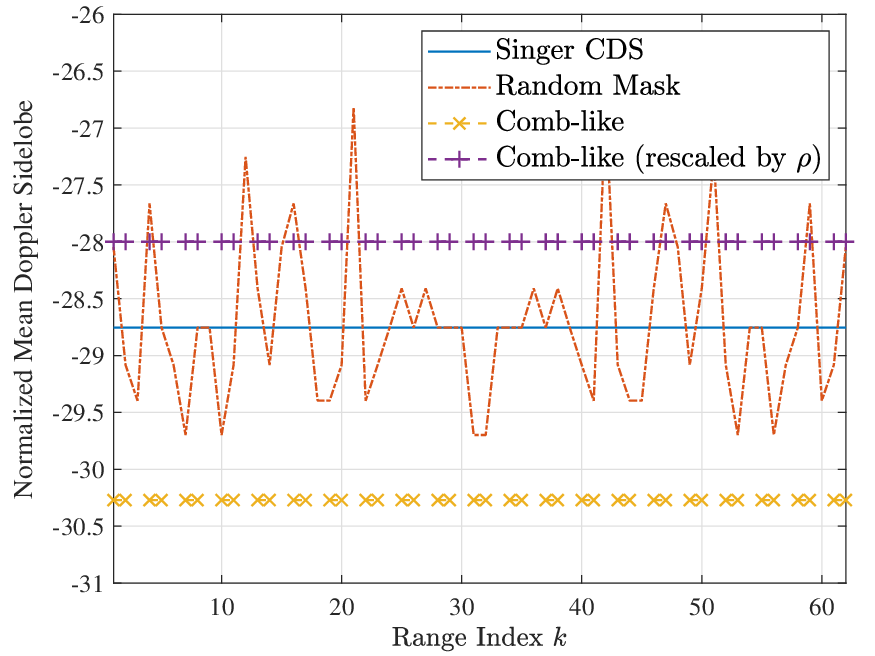}
    \label{fig:average_doppler_sidelobe}
  }
  \caption{The range-Doppler responses of different masks at range mainlobes ($k=l$), full Doppler difference span $\nu\in\{0,\ldots,MN-1\}$.}
  \label{fig:doppler_global}
\end{figure*}

\section{Numerical Results}
\label{sec:num}

In this section, We present numerical evaluations of $\mathbb{E}\{|r(k,l,\nu)|^2\}$ to corroborate the range-Doppler analysis in Sec.~III. In particular, we consider three masks with $N=63$: a Singer \ac{cds} with $\rho=31/63$, a random binary mask with the same $\rho$, and a comb-like mask with $\rho=1/3$. Each mask is repeated by $M=50$ times. We consider 16QAM constellations corresponding to $\mu_4\approx 1.32$.

\subsubsection{Range mainlobe ($k=l$), global Doppler behavior (highly dynamic regime)}
Fig.~\ref{fig:doppler_global} plots $\mathbb{E}\{|r(k,k,\nu)|^2\}$ over the full Doppler difference span $\nu\in\{0,\ldots,MN-1\}$. The periodic masking structure manifests itself as a pronounced, highly structured ambiguity pattern. In particular, the comb-like achieves a considerably low Doppler sidelobe level at most Doppler bins, at the cost of a grating-lobe-like behavior at certain Doppler bins, due to its strict periodicity. By contrast, the Singer \ac{cds} and random masks yield markedly different global distributions. While both exhibit ambiguity induced by periodic operation, the \ac{cds} tends to provide a more favorable global range-Doppler profile as discussed in Sec.~III (e.g., worst-case behavior along the range mainlobe), whereas the random mask shows less controlled variations. We note that the comb-like mask has a smaller $\rho$. According to \eqref{average_sidelobe}, its sidelobe levels would be higher if it also had $\rho=31/63$. Consequently, as portrayed in Fig.~\ref{fig:average_doppler_sidelobe}, the worst case mean Doppler sidelobe level over $k$ of the comb-like mask become larger than that of the \ac{cds}, after rescaling by $\rho$.

\subsubsection{Range mainlobe ($k=l$), zoomed in (moderately dynamic regime)}
\begin{figure*}[t]
  \centering
  \subfloat[Singer \ac{cds}]{
    \includegraphics[width=0.32\textwidth]{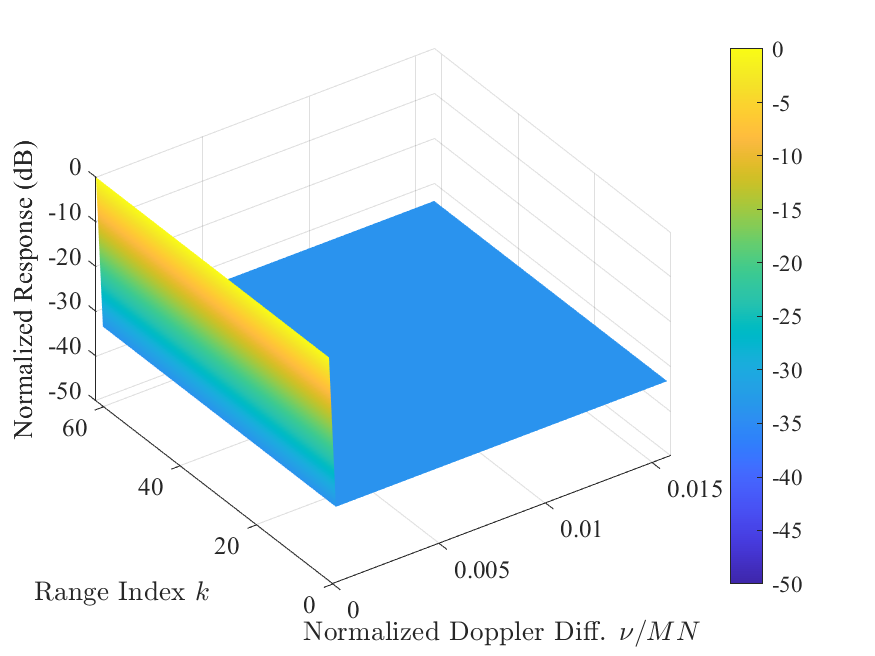}
    \label{fig:doppler_local_cds}
  }
  \subfloat[Random mask]{
    \includegraphics[width=0.32\textwidth]{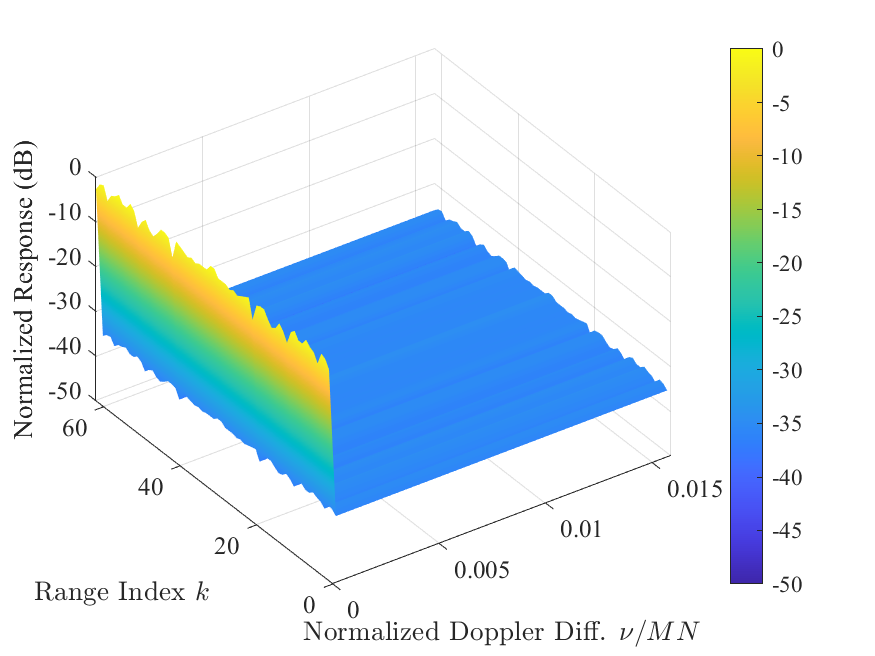}
    \label{fig:doppler_local_rand}
  }
  \subfloat[Comb-like mask]{
    \includegraphics[width=0.32\textwidth]{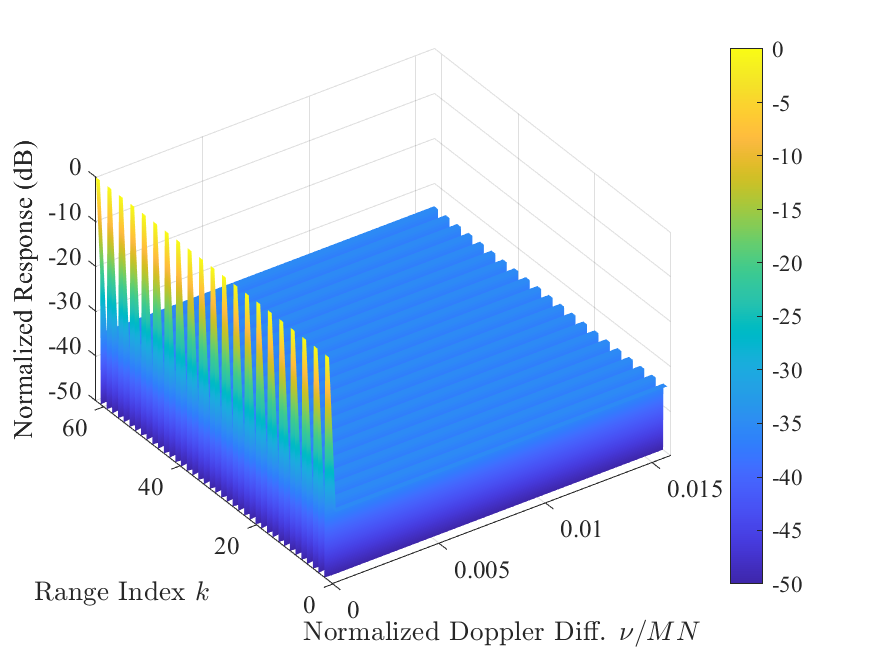}
    \label{fig:doppler_local_comb}
  }
  \caption{The range-Doppler responses of different masks at range mainlobes ($k=l$), local Doppler window $\nu\in\{0,\ldots,M-1\}$ (zoom-in of Fig.~\ref{fig:doppler_global}).}
  \label{fig:doppler_local}
\end{figure*}

Fig.~\ref{fig:doppler_local} zooms into the first Doppler difference block $\nu\in\{0,\ldots,M-1\}$ of Fig.~\ref{fig:doppler_global}, corresponding to the moderately dynamic regime. In this regime, as discussed in Sec.~III, the deterministic Doppler-induced term $|S_{k,MN}(\nu)|^2$ does not contribute to nonzero Doppler bins within this local window, and the remaining behavior is controlled by the mask-dependent factor $a[k]$. Accordingly, masks that equalize $a[k]$ yield smaller and more stable local Doppler sidelobes. The Singer \ac{cds} mask exhibits the most favorable local behavior, which is consistent with its minimax optimality in the moderately dynamic regime, while the random mask shows larger mainlobe fluctuation and higher sidelobe levels. The comb-like mask, although does not exhibit the grating lobes in this local window, suffers from the most severe mainlobe fluctuation.

\begin{figure*}[t]
  \centering
  \subfloat[Singer \ac{cds}]{
    \includegraphics[width=0.32\textwidth]{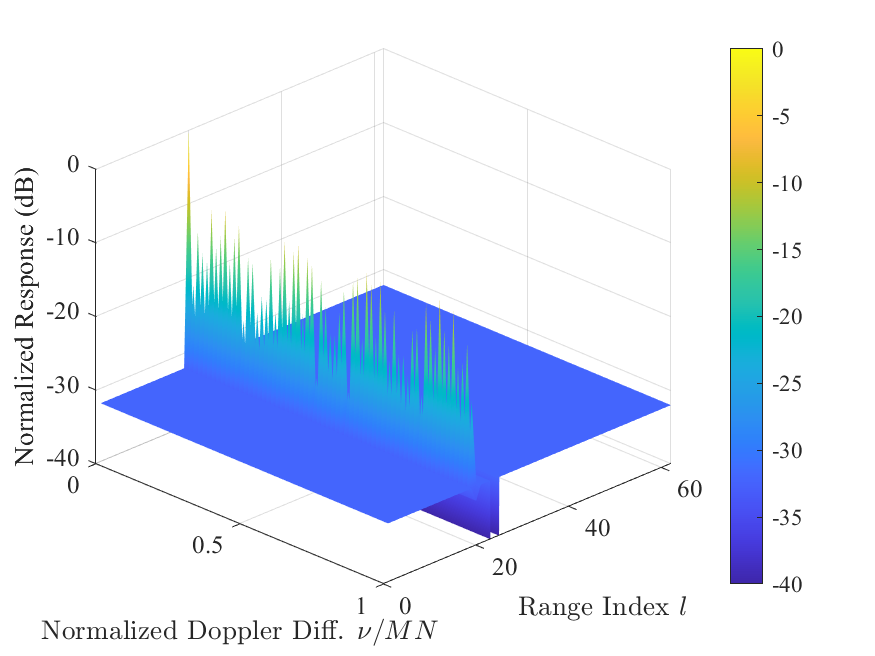}
    \label{fig:range_2d_cds}
  }
  \subfloat[Random mask]{
    \includegraphics[width=0.32\textwidth]{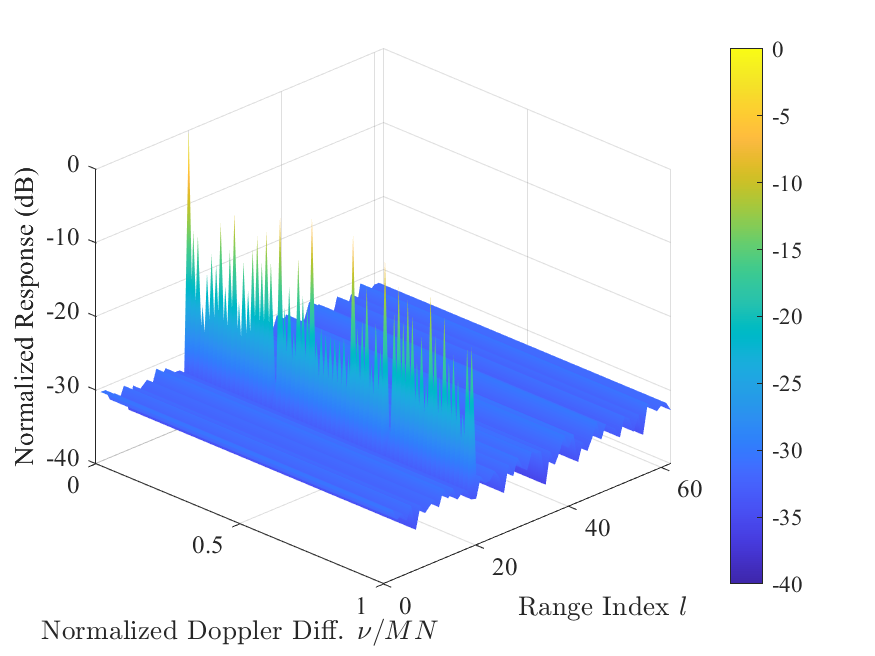}
    \label{fig:range_2d_rand}
  }
  \subfloat[Zero-Doppler-difference slice]{
    \includegraphics[width=0.32\textwidth]{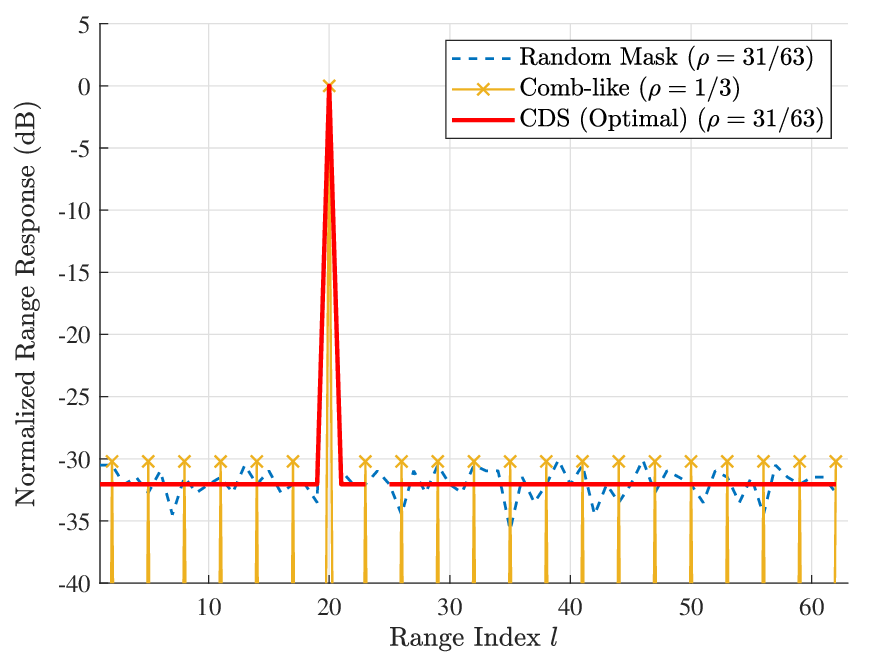}
    \label{fig:range_1d}
  }
  \caption{The range-Doppler response of the Singer \ac{cds} and a random mask, at $k=20$.}
  \label{fig:range_sidelobes}
\end{figure*}

\subsection{Range-Doppler response at a fixed $k$}
Fig.~\ref{fig:range_sidelobes} examines the range-Doppler response at a specific true range bin $k=20$, with varying trial range bin $l$. The two 2D plots show that the sidelobe surface is essentially flat along the Doppler axis, supporting the Doppler-invariance property for $k\neq l$ established in Sec.~III. Consequently, one should evaluate the sensing performance of different masks in the range-sidelobe domain by considering their range-domain correlation structures. This is further highlighted by the zero-Doppler-difference slice portrayed in Fig.~\ref{fig:range_1d}, where the Singer \ac{cds} exhibits the expected sidelobe control advantage over both the random mask and the comb-like mask.

\section{Conclusions}
This paper studied transmission mask design for \ac{masm} in half-duplex \ac{isac} for range-Doppler sensing, by characterizing the expected squared response $\mathbb{E}\{|r(k,l,\nu)|^2\}$. We showed that range sidelobes ($k\neq l$) are Doppler-invariant, enabling direct extension of the range-sidelobe optimality in the ranging-only scenario. For the range mainlobe ($k=l$), periodic masking induces a sparse Doppler-sidelobe structure. Consequently, in the moderately dynamic regime, Singer \acp{cds} are minimax-optimal, simultaneously suppressing worst-case Doppler sidelobes and reducing mainlobe fluctuation. By contrast, in the highly dynamic regime, we quantified the aggregate deterministic Doppler-sidelobe energy along the range mainlobe and showed it is a concave function of the mask autocorrelation, revealing an inherent tradeoff between minimizing mainlobe fluctuation and minimizing Doppler sidelobes.

Our results provide explicit guidance on when \ac{cds}-type masks remain ideal and when tradeoffs are unavoidable, and motivate future work on ambiguity-resolving strategies and mask design under additional practical constraints.

\appendices
\section{Proof of Proposition \ref{prop:doppler_sidelobe_range_main}}\label{sec:proof_doppler_sidelobe_range_main}
\begin{IEEEproof}
Note that for an integer $\nu$, the term $S_{k,N}(\nu)$ is the discrete Fourier transform of $\gamma_k[n]:=m_{\rm r}[n]m_{\rm t}[n-k]$. According to Parseval's identity, we have
\begin{align}
\sum_{\nu=1}^{N-1}|S_{k,N}(\nu)|^2&=\sum_{\nu=0}^{N-1}|S_{k,N}(\nu)|^2 - |S_k(0)|^2 \nonumber\\
&= N\sum_{n=0}^{N-1}|\gamma_k[n]|^2-|S_k(0)|^2\nonumber\\
&=N\sum_{n=0}^{N-1}|\gamma_k[n]|^2-\left(\sum_{m=0}^{N-1}\gamma_k[m]\right)^2. \label{parseval}
\end{align}
Next, note that $|\gamma_k[n]|^2=\gamma_k[n]$ since $\gamma_k[n]$ is a $0/1$ sequence. We may thus obtain
$$
\begin{aligned}
\sum_{n=0}^{N-1}|\gamma_k[n]|^2&=\sum_{n=0}^{N-1} (1-m_{\rm t}[n])m_{\rm t}[n-k]\\
&=\sum_{n=0}^{N-1}m_{\rm t}[n-k]-\sum_{m=0}^{N-1}m_{\rm t}[m]m_{\rm t}[m-k],
\end{aligned}
$$
where we have
$$
\sum_{n=0}^{N-1}m_{\rm t}[n-k] = \rho N,
$$
since cyclic shift does not change the number of non-zero entries. Now, by the definition of $a[k]$, we have
\begin{equation}\label{squared_corr}
\sum_{n=0}^{N-1}|\gamma_k[n]|^2 = \rho N - a[k].
\end{equation}
Substituting \eqref{squared_corr} into \eqref{parseval}, we have
$$
\sum_{\nu=1}^{N-1} |S_{k,N}(\nu)|^2 = (\rho N-a[k])[N-(\rho N-a[k])],
$$
completing the proof.
\end{IEEEproof}

\section{Proof of Proposition \ref{prop:eisl_max}}\label{sec:proof_eisl_max}
\begin{IEEEproof}
We first note that
\begin{equation}
I_{a[k]}=\sum_{k=1}^{N-1} \sum_{\nu=1}^{N-1} |S_{k,N}(\nu)|^2+(\mu_4-1)(\rho N-a[k]).
\end{equation}
Now, observe that
\begin{subequations}
\begin{align}
\sum_{k=1}^{N-1} a[k] &= \sum_{k=1}^{N-1}\sum_{n=0}^{N-1} m_{\rm t}[n]m_{\rm t}[n-k] \\
&=\sum_{n=0}^{N-1}m_{\rm t}[n]\sum_{k=1}^{N-1} m_{\rm t}[n-k]\\
&=\sum_{n=0}^{N-1}m_{\rm t}[n](\rho N-m_{\rm t}[n])\\
&=\sum_{n=0}^{N-1}\rho N m_{\rm t}[n]-(m_{\rm t}[n])^2 \\
&=\rho N(\rho N-1), \label{from_zero_one}
\end{align}
\end{subequations}
where \eqref{from_zero_one} follows from the fact that $m_{\rm t}[n]=(m_{\rm t}[n])^2$. We see that $\sum_{k=1}^{N-1}(\mu_4-1)(\rho N-a[k])$ does not depend on the specific pattern of $a[k]$. Thus using Jensen's inequality, we have
$$
\begin{aligned}
I_{a[k]} &= \sum_{k=1}^{N-1} [f(a[k]) + (\mu_4-1)(\rho N-a[k])] \\
&\leq (N-1)\left[f\left(\frac{\sum_{k=1}^{N-1}a[k]}{N-1}\right)+ (\mu_4-1)\rho (1-\rho)N^2\right] \\
&=(N-1)\bigg[\left(N-\rho N + \frac{\rho N(\rho N-1)}{N-1}\right)\\
&\hspace{3mm}\cdot \left(\rho N-\frac{\rho N(\rho N-1)}{N-1}\right)+(\mu_4-1)\rho (1-\rho)N^2\bigg],\\
&=\rho(1-\rho) N^2\left(N-\frac{\rho(1-\rho) N^2}{N-1}+(N-1)(\mu_4-1)\right)
\end{aligned}
$$
where the equality is achieved when all $a[k]$'s equal to $\frac{\sum_{k=1}^{N-1} a[k]}{N-1}$, and hence the proof is completed.
\end{IEEEproof}

\section{Proof of Proposition \ref{prop:doppler_lower_bound}}\label{sec:proof_doppler_lower_bound}
\begin{IEEEproof}
Since $f(a[k])$ is a concave function of $a[k]$, the summation $I_{a[k]}$ achieves its minimum when the values of $a[k]$ polarize, in the sense that there are $(\rho N-1)$ different values of $k$ where $a[k] = \rho N$, while for other values of $k$ (note that $k\neq 0$) we have $a[k]=0$. In this case, according to \eqref{s_nu}, $I_{a[k]}$ can be written as
\begin{align}
I_{a[k]} &= (\rho N-1) \left(\rho N-\rho N\right)[N-(\rho N-\rho N)] \nonumber \\
&\hspace{3mm} + (N-\rho N) \rho N (N-\rho N) + (\mu_4-1)\rho (1-\rho)N^2 \nonumber \\
&= \rho (1-\rho)^2N^3 + (\mu_4-1)\rho (1-\rho)N^2,
\end{align}
yielding \eqref{doppler_lower_bound}.

In general, such minimum is not necessarily achievable, since there might not exist such a transmission mask $m_{\rm t}[n]$ satisfying the conditions on its autocorrelation sequence $a[k]$. Nevertheless, when $d=1/\rho$ is an integer, there exists comb-like masks satisfying \eqref{comb_seq}. For these masks, if $k$ is an integer multiple of $d$, say $k=md,~m=1,\dotsc,\rho N-1$, for any $n=ld,~l=0,\dotsc,\rho N-1$ at which $m_{\rm t}[n]=1$, we have $n-k=(l-m)d$. It then follows that $(l-m)d~{\rm mod}~N$ is also an integer multiple of $d$. As $l$ runs over $0,\dotsc,\rho N-1$, there are exactly $\rho N$ values of $l$ satisfying $m_{\rm t}[n-k]=1$, and hence we have $a[k]=\rho N$. Since $k=md,~m=1,\dotsc,\rho N-1$, there are $\rho N-1$ such values of $k$.

On the other hand, when $k$ is not an integer multiple of $d$, for any $n=ld,~l=0,\dotsc,\rho N-1$ at which $m_{\rm t}[n]=1$, $n-k$ has to be an integer multiple of $d$ for $m_{\rm t}[n-k]$ to be $1$, but this is impossible since $d$ is not a divisor of $k$. Thus for any $k$ that is not an integer multiple of $d$, we must have that $m_{\rm t}[n-k]=0$. Therefore, we may now conclude that the autocorrelation sequences $a[k]$ of comb-like sequences indeed have $\rho N-1$ entries taking the value of $\rho N$, while others $(k\neq 0)$ take the value of $0$, and hence attain the equality of \eqref{doppler_lower_bound}. Thus the proof is completed.
\end{IEEEproof}

\bibliographystyle{IEEEtran}
\bibliography{references_ieeeabrv_titlebrace}
\end{document}